# Seismic data interpolation based on U-net with texture loss

Wenqian Fang[1], Lihua Fu[1], Meng Zhang[2], Zhiming Li[1*]

[1]School of Mathematics and Physics, China University of Geosciences, Wuhan 430074, China;

[2]Department of Computer Science, Central China Normal University, Wuhan 430079, China

**ABSTRACT:** Missing traces in acquired seismic data is a common occurrence during the collection of seismic data. Deep neural network (DNN) has shown considerable promise in restoring incomplete seismic data. However, several DNN-based approaches ignore the specific characteristics of seismic data itself, and only focus on reducing the difference between the recovered and the original signals. In this study, a novel Seismic U-net InterpolaTor (SUIT) is proposed to preserve the seismic texture information while reconstructing the missing traces. Aside from minimizing the reconstruction error, SUIT enhances the texture consistency between the recovery and the original completely seismic data, by designing a pre-trained U-Net to extract the texture information. The experiments show that our method outperforms the classic state-of-art methods in terms of robustness.

## 1 Introduction

Variable operating conditions during seismic surveys frequently result in inadequate trace spacing along spatial axes. The absence of data affects subsequent offset imaging, inversion, and interpretation, as well as the description of geological structures. Therefore, seismic data interpolation is essential in seismic surveys.

Seismic data reconstruction techniques can be divided into four categories: prediction filters, wave equations, transform domains, and low-rank theory. Prediction-filter-based methods involve the convolution of seismic data with filters, which mainly include the *f-x* domain seismic traces interpolation method (Spitz, 1991) and the t-x domain method (Claerbout and Nichols, 1991). Methods based on wave-equations typically use wave propagation to reconstruct seismic data via iterative solution of forward operators and inversion operators (Bagaini and Spagnolini, 1993; Ronen, 1987). Transform-based approaches generally seek 'optimal' coefficients in transform domain in terms of least squares; and a good reconstruction can be obtained via inverse transformation. These approaches include Radon transform (Thorson and Claerbout, 1985), Fourier transform (Duijindam et al., 1999; Liu and Sacchi, 2004; Zwartjes and Sacchi, 2007), and Curvelet transform (Herrmann and Hennenfent, 2008; Trickett et al., 2010). Rank-reduction-based methods are based on the assumption that seismic data with a limited number of events are of low rank in the *f-x* domain, and the missing data and random noise will increase the rank of the matrix or tensor. Therefore, rank-reduction schemes have become a popular tool for seismic data reconstruction (Gao et al., 2013; Naghizadeh and Sacchi, 2012; Kreimer, 2013; Gao et al., 2017).

Deep learning (DL) establishes deep neural networks that simulate the human brain for analysis and learning. It has been applied successfully to speech recognition, facial recognition, video classification, and texture recognition. In recent years, DL has attracted increasing attention for addressing the problems pertaining to seismic data interpolation. Wang et al. (2019) used a residual network (He et al., 2016) pre-interpolated by bicubic for seismic data antialiasing interpolation. This approach was demonstrated to achieve better results than the *f-x* method; however, it was also recognized that the interpolation bias increased as the feature differences between the test dataset and the training dataset increased. Dario et al. (2018) used a conditional generative adversarial network for the interpolation problem in post-stack seismic datasets. They established a network pool for different gap widths of missing traces; this approach was shown to be better than a single network in terms of the Pearson correlation coefficient. However, the limitation of this network pool method is that it requires considerable amounts of data and calculations. Mandelli et al. (2018) applied the U-net network to the random missing interpolation problem of seismic data and they achieved better results than the classical low-rank Singular Spectrum Analysis (SSA) algorithm.

Although the DL method has considerable potential in the field of seismic data interpolation, it is also characterized by certain limitations. (1) Majority of the existing methods refer to the results in computer vision but do not focus on the differences in the seismic data. (2) To overcome the challenge of generalizing the learned knowledge to new datasets, a large number of training samples are required. However, seismic exploration is an expensive venture, and seismic data set for deep learning algorithms like the ImageNet dataset (Russakovsky, 2015) in computer vision, which contains rich seismic data features, is lacking. Therefore, the generalization issue has become more prominent. In the present study, we propose an algorithm, the Seismic U-net InterpolaTor (SUIT) algorithm, to address texture loss for seismic data; good generalization considering only a few training samples is achieved.

We use a well-known architecture U-net (Olaf et al., 2015) as the backbone of SUIT. It was originally designed for addressing image segmentation problems and has become widely used in computer vision and image processing such as object detection (Lin et al., 2017), image translation (Isola, 2017), image denoising (Heinrich, 2018) and image inpainting (Pathak, 2016). Moreover, it can work with very few training samples (Olaf, 2015).

In the area of seismic data interpolation, Mandelli et al. (2018) applied U-net to seismic data interpolation and achieved satisfactory results. However, this method still performs a poor generalization; their further research showed that the interpolation bias increased noticeably in case of transfer learning (Mandelli et al., 2019). In our experiments, we find that the interpolated data often manifests itself as lack of texture in the missing regions. The problem often occurs on the test data when presence of dense parallel events or multiple consecutive traces are missing. In the case of the presence of dense parallel events, the prediction of one event is influenced by the others as the events are similar and difficult to distinguish. Besides, because of convolution architecture, the reconstruction may be obviously influenced by the initial hole values (always zero) when multiple consecutive traces are missing. Example of transfer learning using a U-net architecture can be seen in Figure 1(c). The details of false events are highlighted by the circle. Figure 1(d) shows the result of our method in the same situation, where the events are more continuous.

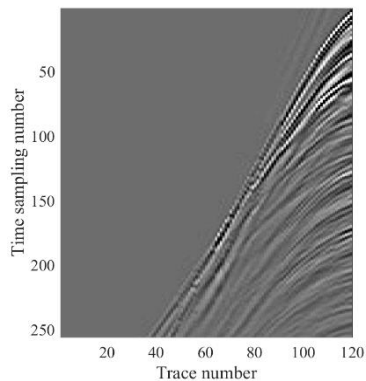
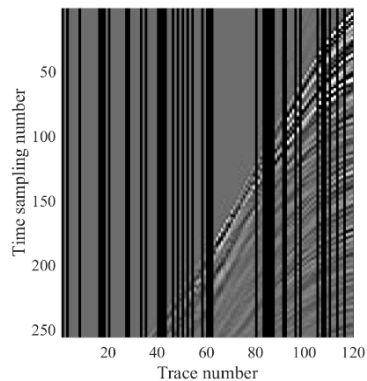

(a) orignal data                (b) sampled data

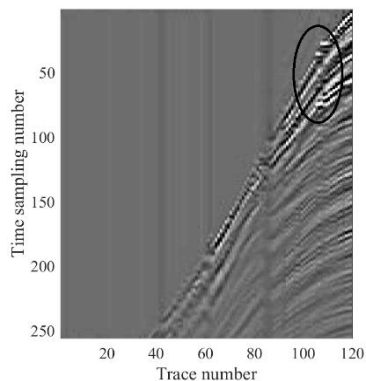
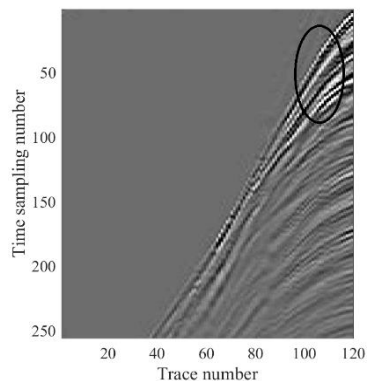

(c) classic U-net          (d) our method

Figure 1. Comparison between our algorithm and classic U-net

We propose the texture loss that enhances the texture consistency between the original and the recovered seismic records. As the recorded seismic signal consists of amplitudes of various strength and sign (peaks or troughs), we define the texture of the seismic data as the peaks and troughs which can represent the boundaries between rock layers. First, we apply K-means to identify these pixels. Then, the segmentation is considered for labels to train the texture extractor. In this manner, the texture loss information can be propagated backward to make a good effect on Deep neural network (DNN) parameters training. The SUIT algorithm framework proposed herein is presented in Fig. 2; it consists of two networks cascade: a reconstructed backbone and a texture extractor and both networks use U-net. After training, only the reconstructed network is used to recover the signal and the computational costs will not increase.

We evaluated our method against the classic U-net and the well-known Low-rank Matrix Fitting (Wen et al., 2012) on post-stack and pre-stack data. The experiments show that our method has better robustness, and the recovered signals are more continuous along the spatial direction with good texture details.

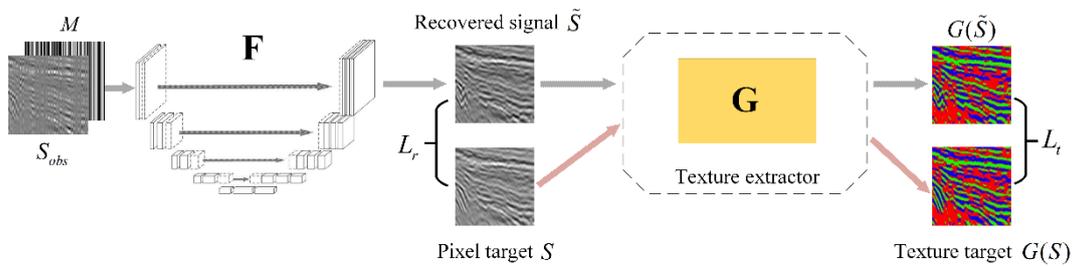

Figure 2. SUIT algorithm framework. Network $F$ is used to reconstruct the data and network $G$ is used to extract texture features.

## 2 Method

### 2.1 U-net architecture

We use classic U-net as the backbone for the reconstructed network as well as for the texture extractor. Figure 3 shows the U-net architecture; it consists of downsampling (left side) and upsampling (right side) processes. At each downsampling phase (represented by black downward arrows in Fig. 3), maximum pooling is applied to half the feature map in height and width, but the number of channels remain unchanged. The upsampling process uses deconvolution (as shown by the white upward arrows in the figure) that doubles the height and width of the feature map and the number of channels is halved. The skip connection is the most distinctive structure of the U-net network, which is known to produce detailed output. The feature map in the downsampling process (shown by the dashed boxes) is directly concatenated with the corresponding upsampled feature map. More details regarding U-net are present in Olaf et al. (2014).

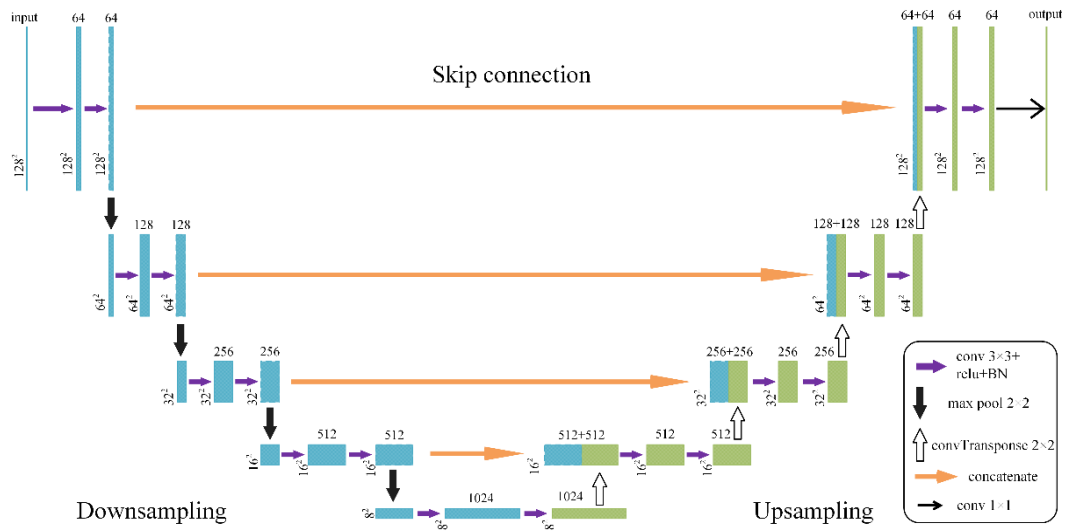

Figure 3. Structure of the U-net network.

### 2.2 SUIT design for interpolation

The original problem of seismic data interpolation can be formulated as $S_{obs} = MS$, where $S_{obs}$ is the observation data with traces missing, $S$ is the complete data, and $M$ is the sampling matrix (mask operator); the goal is to estimate $S$ from $S_{obs}$ and $M$. DL-based methods have shown great potential in seismic interpolation, for they extract high-level semantic information hidden behind training data in a nonlinear manner to provide reliable estimates of complete data. In this paper, we design SUIT for seismic interpolation; the architecture is shown in Figure 2. This model consists of two U-net networks, $F$ and $G$. We use network $F$ to reconstruct the seismic data, and the interpolation inference can be represented as $\tilde{S} = F(S_{obs}, M, \theta_F)$, where $\tilde{S}$ denote the recovered signal and $\theta_F$ denotes the parameter of $F$. Network $G$ is designed to propagate the texture information of seismic data; the details are discussed in subsequent sections.

**Texture Segmentation.** It is well known that seismic data have rich texture information, which is an expression of the stratigraphic configuration of beds and thin beds. For seismic restoration, we defined the texture of the seismic data as peaks and troughs reflecting information of different horizons. We used the image segmentation algorithm K-means to divide the signal into three parts, that are presented as RGB three-channel color images in Fig. 4. In Fig. 4b and 4d, green, blue and red colors indicate peak, trough, and background regions, respectively. Enhanced precision in segmentation enables more accurate texture extraction; however, the experimental section shows that even approximate results from the K-means have significant effect on the ultimate reconstruction.

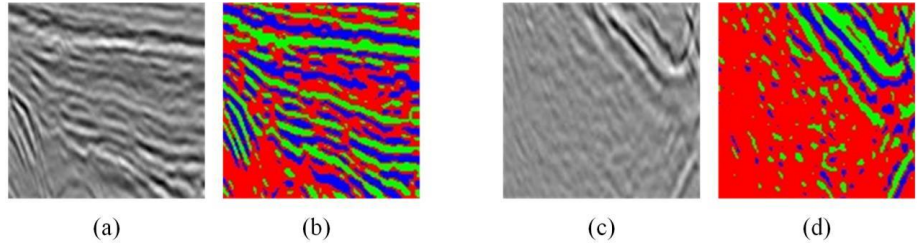

Figure 4. Texture of seismic data by K-means algorithm: (a) and (c) are seismic data, (b) and (d) are texture information

**Texture Extractor *G*.** As shown in Fig.2, the network *G* is pre-trained to identify the texture segmentation from the feeding seismic image. The input of *G* is the complete seismic record and the output is the corresponding texture segmentation (see Fig. 5). As a 2-D classification task, we selected the cross-entropy between the output and the texture segmentation from K-means as the loss function:

$$L_G(\theta_G) = -\sum_{x \in \Omega} \log(G_{l(x)}^x(S, \theta_G)) \quad (1)$$

where $\theta_G$ represents the parameters of *G*; $x \in \Omega, \Omega \subset \mathbb{Z}^2$ represents the position coordinates of the data, the function $l: \Omega \to \{1, 2, \ldots, K\}$ represents the true category of each point *x*, and $G_{l(x)}^x(S, \theta_G) \in (0,1)$ is the output of *G* (the softmax function converts the output to a classification probability), which indicates the probability that *x* belongs to the *k*-th class. Optimizing $\theta_G$ suggests the probability that point *x* is classified to the real category $l(x)$ close to 1.

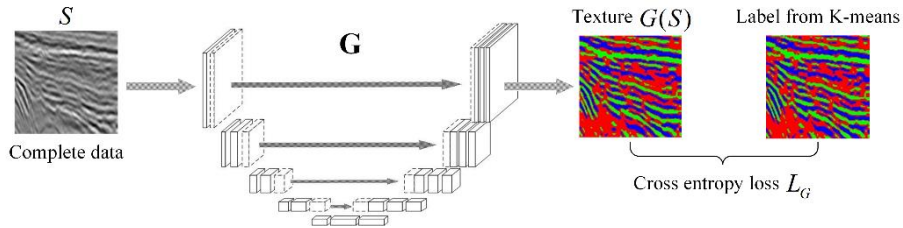

Figure 5. Training of texture extractor.

**Reconstruction Network *F*.** Network *F* is designed for the reconstruction task. As can be seen in

Fig. 2, the input of F is the stacking of random sampled data $S_{obs}$ and mask M, and the output is the recovered signal $\tilde{S}$. The loss function of F, targets both per-pixel reconstruction accuracy as well as texture consistency between the original and approximated seismic signals.

For the pixel constraint, the reconstruction loss $L_r$ is defined as the Frobenius norm of the difference between recovered and original signals:

$$L_r(\theta_F) = \left\| \tilde{S} - S \right\|_F^2 \tag{2}$$

Here $\tilde{S}$ is the output of reconstruction from network F, while $S$ denotes the original data. Minimizing $L_r$ result in the output being as close as possible to the complete data.

The reconstruction loss $L_r$ is one of the most popular loss functions used in image reconstruction. Many image restoration solutions modify $L_r$ by applying various constraints to promote DNN to infer more realistic and reliable results. For example, perceptual loss is used to reduce perceptual differences (Johnson et al., 2016), total variation loss helps enhance the reconstructed image smoothness (Liu et al., 2018), and style loss produces the image with desired style (Iizuka et al., 2017). Based on these, we designed texture loss $L_t$ to reduce the texture difference between the original and approximated seismic signal. To the best of our knowledge, no research has introduced texture loss into seismic restoration.

$L_t$ is defined as the Frobenius norm between the texture of recovered and original signals:

$$L_t(\theta_F, \theta_G) = \left\| G(\tilde{S}) - G(S) \right\|_F^2 \tag{3}$$

Here $G(\tilde{S})$ and $G(S)$ are outputs of G, which represent the texture of reconstruction $\tilde{S}$ and original signal $S$ respectively.

Finally, the total loss $L_F$ is the combination of $L_r$ and $L_t$, as shown in Equation (4),

$$L_F(\theta_F, \theta_G) = \lambda L_r(\theta_F) + L_t(\theta_F, \theta_G) \tag{4}$$

where $\lambda$ indicates the weight (the larger the magnitude of $\lambda$, the larger the weight of $L_r$). $L_F$ is used to optimize the network $F$, enforcing consistency of pixels and textures simultaneously.

**2.3 Pipeline of SUIT**

The pipeline of SUIT mainly contains the K-means segmentation, training of network $G$ and training of network $F$, details are as follows:

---

**Algorithm** SUIT for seismic data interpolation

The algorithm receives the sampled data $S_{obs}$, sampling matrix $M$, complete signal $S$, and randomly initialized parameters $\theta_F$, $\theta_G$ of network $F$ and $G$, respectively, as input. Let $l(S)$ be the texture segmentations of $S$ by K-means. The first stage is a cross-entropy training of network $G$ (Step 2). The second stage is the training of network $F$, based on the prediction error of pixel and texture (Step 3). Finally, the output $\tilde{S}$ is the inference of $F$.

**Input:** $S_{obs}, M, S, \theta_F, \theta_G$

**Output:** $\tilde{S}$

Step1: Obtain the texture segmentations $l(S)$ by K-means

Step2: Train texture extractor $G$: $\theta_G = \arg\min_{\theta_G} L_G(\theta_G)$

Step3: Train $F$: $\theta_F = \arg\min_{\theta_F} L_F(\theta_F, \theta_G)$

Step4: Obtain output $\tilde{S} = F(S_{obs}, M, \theta_F)$

---

**3 Experiment**

**3.1 Network's training**

Our experiment involves post-stack data as a training dataset. As shown in Fig. 6a, this data has 2501 traces and each trace has 1011 temporal sampling points. We cropped the dataset to $128 \times 128$ size with 50% overlap to obtain 782 elements as a training dataset $\Omega:\{S_1, S_2, ..., S_{782}\}$. To evaluate a network with improved generalization performance, the verification set use another work area post-stack data (Fig. 6b). It has 512 traces and each trace has 512 temporal sampling points. Similarly, we cropped it into 109 128 × 128 blocks for verification.

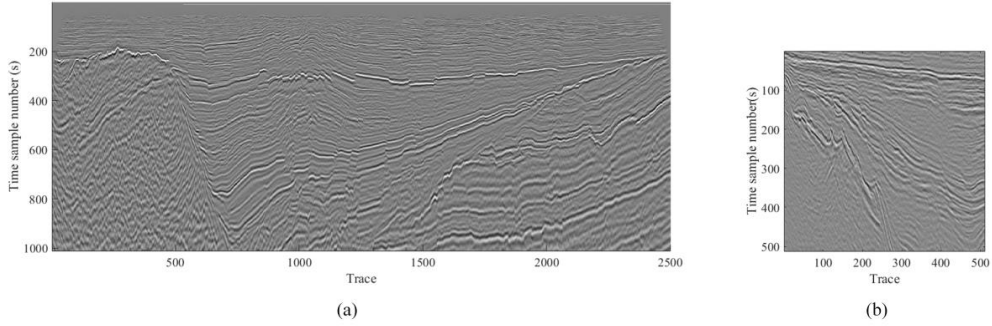

Figure 6. Training and verification datasets: (a) training dataset and (b) verification dataset.

Because the data set was very small, to prevent overfitting, we first performed a random left–right flip and then randomly dismissed 50% of the traces in each iteration. Consequently, for one element of data, the position of missing data was different in each iteration. The data were trained for 150 epochs. Training time was approximately 1 h for two Nvidia 1080ti graphics processing units. The model was evaluated based on the average Signal-to-Noise Ratio (SNR) of the verification set after each epoch, as shown in Equation (5):

$$S/R(dB) = 10\log_{10}\frac{\|S\|_2^2}{\|S - \tilde{S}\|_2^2} \qquad (5)$$

where $S$ is the label and $\tilde{S}$ is the recovered data. Figure 7 shows the average SNR of the verification set at every epoch during the training step. The dotted line shown in Fig. 7 represents the U-net results and the solid line illustrates the SUIT results. It can be seen that as the iteration progresses, the SNRs of the two algorithms gradually increase and after 100 epochs they approach convergence. Ultimately, we selected the model with the highest SNR as the final model. The highest SNR of the verification set for U-net model was 24.74 dB, while the SNR of the SUIT model was 27.76 dB, i.e., approximately 3 dB higher.

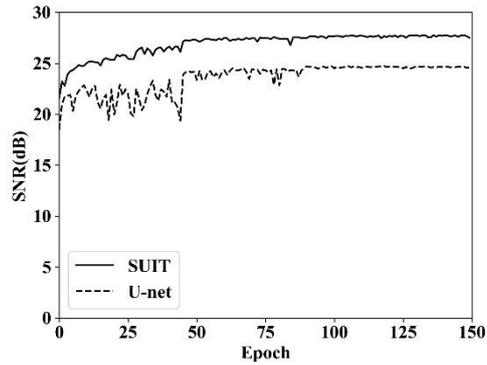

Figure 7. Reconstruction SNR of the verification set during the training step.

We monitored the reconstruction of one verification dataset under different epochs, as shown in Fig. 8. The behavior illustrated in the upper panels of the figure shows texture loss, i.e., reconstruction by the SUIT model, while the behavior in the lower panels shows no texture loss, i.e., from the U-net model. It can be seen from the marked area that the SUIT model is more accurate in reconstructing texture as the number of iterations increases, whereas the texture of the U-net model exhibits less continuity.

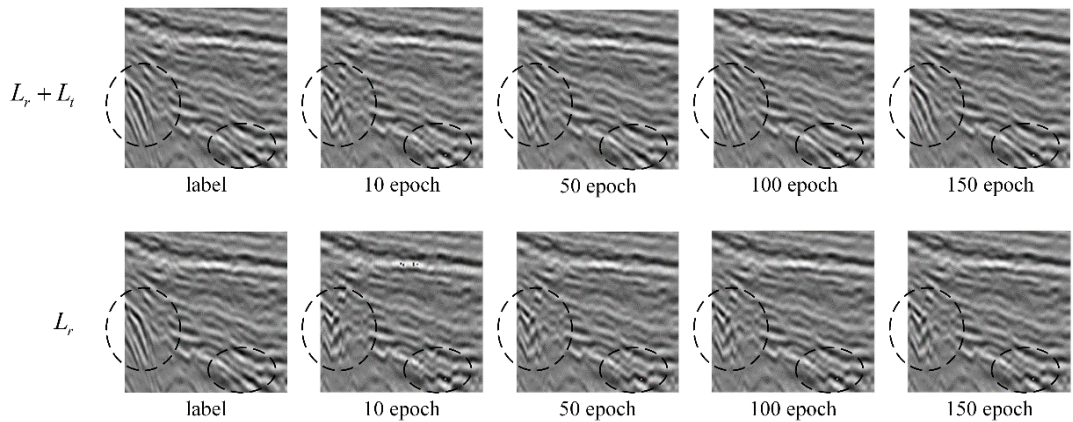

Figure 8. Comparison between the SUIT and U-net models in terms of texture accuracy.

**3.2 Post-stack data interpolation**

Using the trained network $F$ in the SUIT, trained U-net, and LMaFit algorithms we performed random missing interpolation on the field post-stack data $S_{post}$ used in the verification dataset. These data consist of 512 traces with 512 temporal samples per trace, as shown in Fig. 9a. Figure 9b shows the sampled data with 50% data missing, Fig. 9c–h shows the interpolation and the corresponding residuals of the SUIT algorithm, U-net model, and LMaFit algorithm. The interpolation SNR of the SUIT algorithm is 33.47 dB, while that of the U-net and LMaFit methods is 29.93 and 23.77 dB, respectively. The dashed box shown in the figure is a partially enlarged view of the solid-line frame area. It can be seen that the reconstruction by the DL method is smoother than that obtained by the LMaFit algorithm. Moreover, the SUIT method reconstructs the texture details more accurately than the other methods, which can better recover the original appearance of the seismic data.

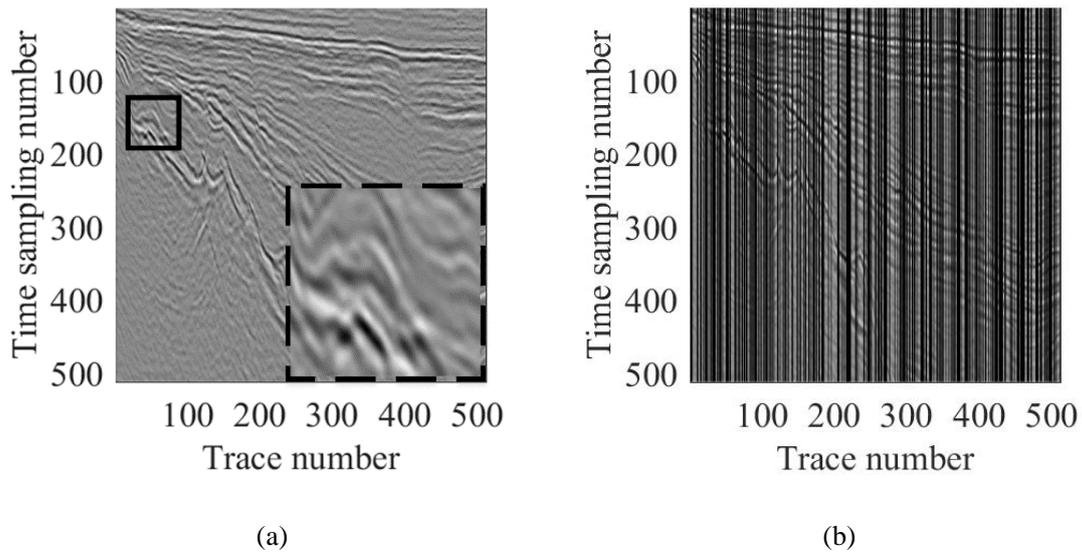

(a)          (b)

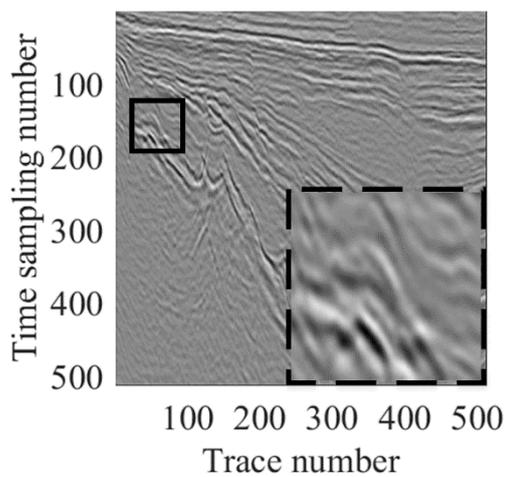
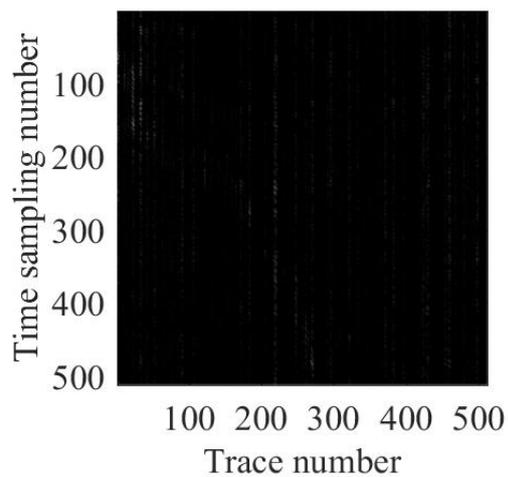

(c)　　　　　　　　　　　　　　(d)

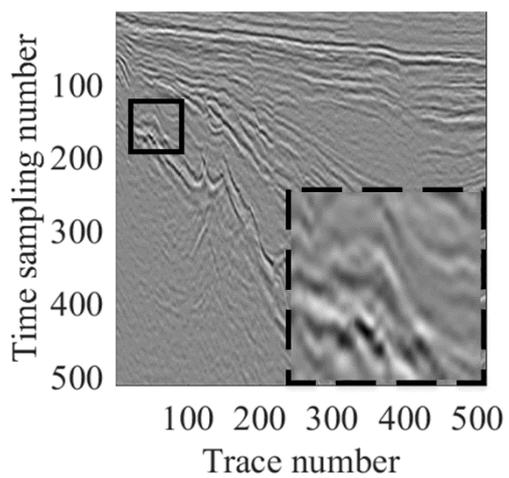
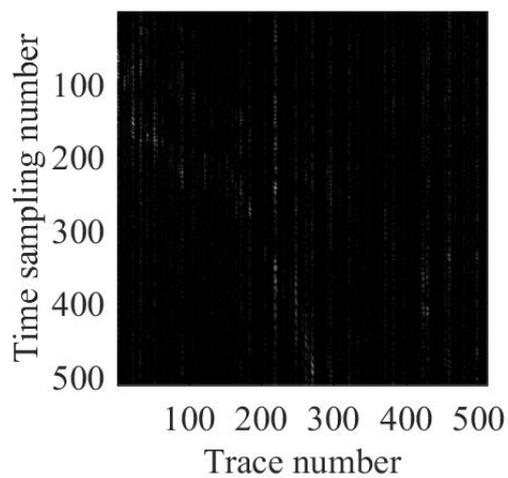

(e)　　　　　　　　　　　　　　(f)

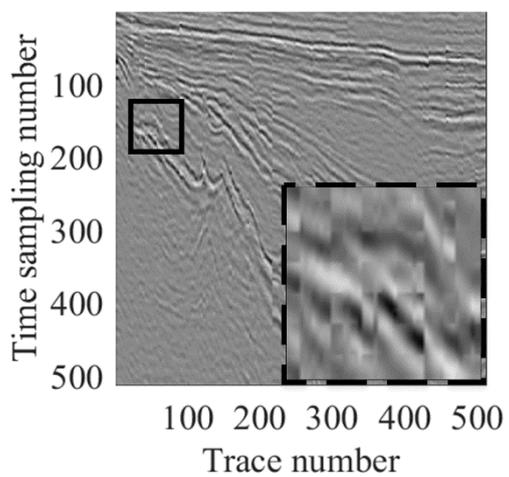
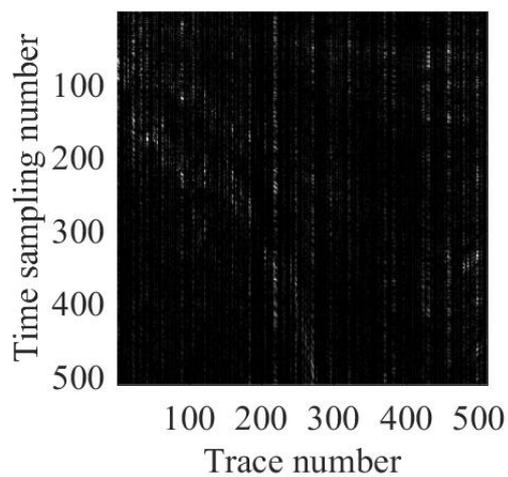

(g)　　　　　　　　　　　　　　(h)

Figure 9. Interpolation comparisons of post-stack data: (a) complete data; (b) random sampled data with 50% data missing; (c) and (d) interpolated data and the residual, respectively, using the trained network *F* in the SUIT model; (e) and (f) interpolated data and the residual, respectively, using the trained U-net model; and (g) and (h) interpolated data and the residual, respectively, using the LMaFit method

Figure 10 shows the corresponding *f-k* spectra of Fig. 9, where the *x*-axis represents the normalized wave number and the *y*-axis represents frequency. It can be seen from the figure that the reconstruction of the DL method is more concentrated and has better anti-aliasing ability, while the reconstruction of the LMaFit algorithm still exhibits the dispersion phenomenon.

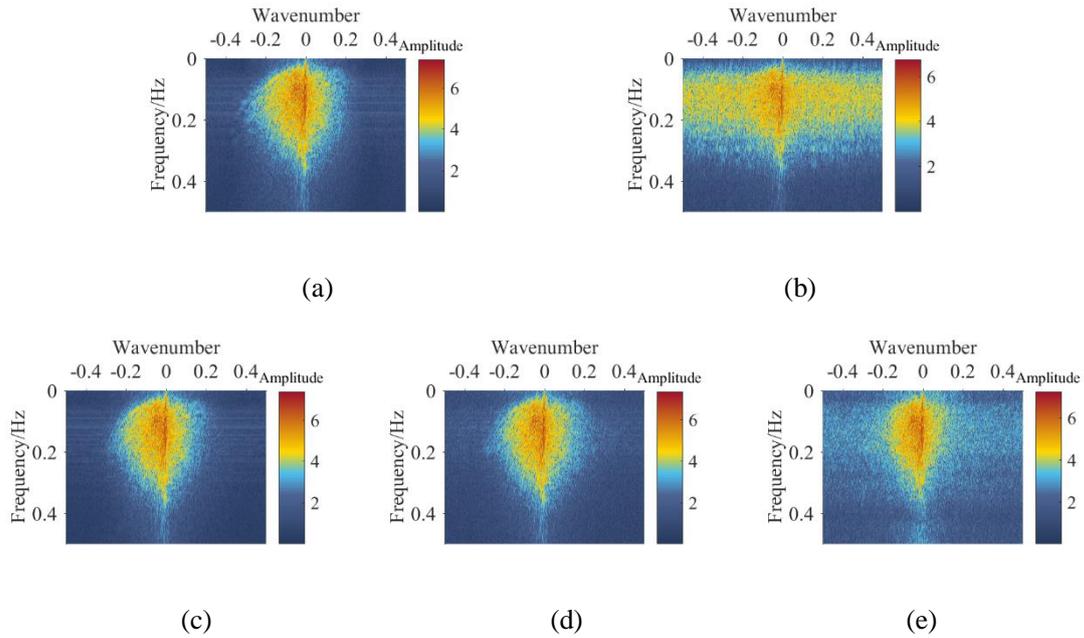

(a) (b)

(c) (d) (e)

Figure 10. Comparison of the f-k spectra of post-stack data: (a) complete data, (b) sampled data with random 50% missing, (c) interpolated data using the trained network *F* in the SUIT algorithm, (d) interpolated data using the trained U-net model, and (e) interpolated data using the LMaFit method.

Figure 11 shows the results of the single-trace reconstruction of the 270th trace of the post-stack data. Figure 11a, 11c, and 11e are the single-channel reconstruction of the trained $F$ in the SUIT algorithm, trained U-net model, and LMaFit method, respectively, where the blue line represents the original data and the red line represents the reconstruction. The horizontal axis represents the amplitude and the vertical axis represents the temporal sampling point. (Figure 11b, 11d, and 11f are the corresponding residual maps.) It can be seen that the SUIT method can fit the single trace signal well and that the reconstruction error is small. The interpolated data using the U-net model has certain error in the peak region, which is more obvious in the area indicated by the arrow. Comparison of the residuals reveals that the interpolated data of the SUIT method have a smaller single-trace reconstruction error.

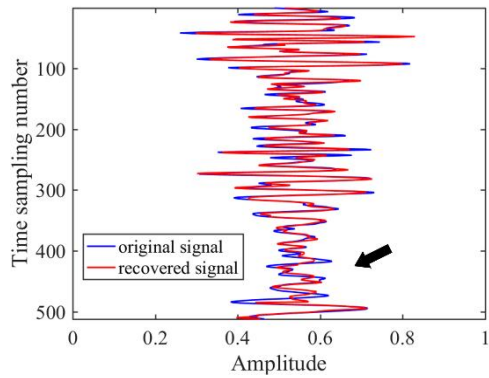

(a)

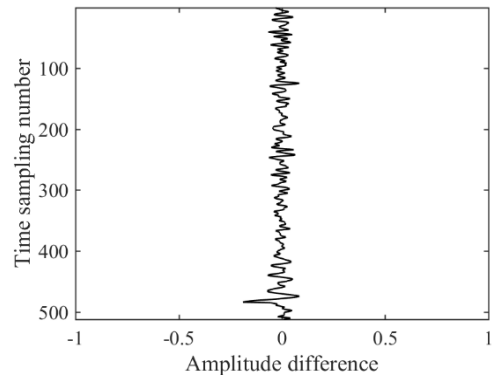

(b)

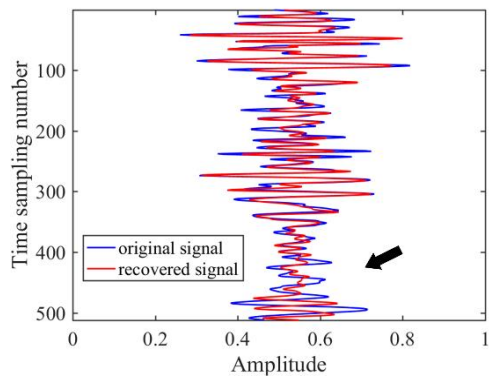

(c)

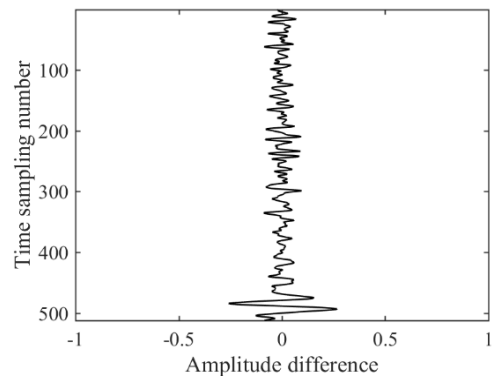

(d)

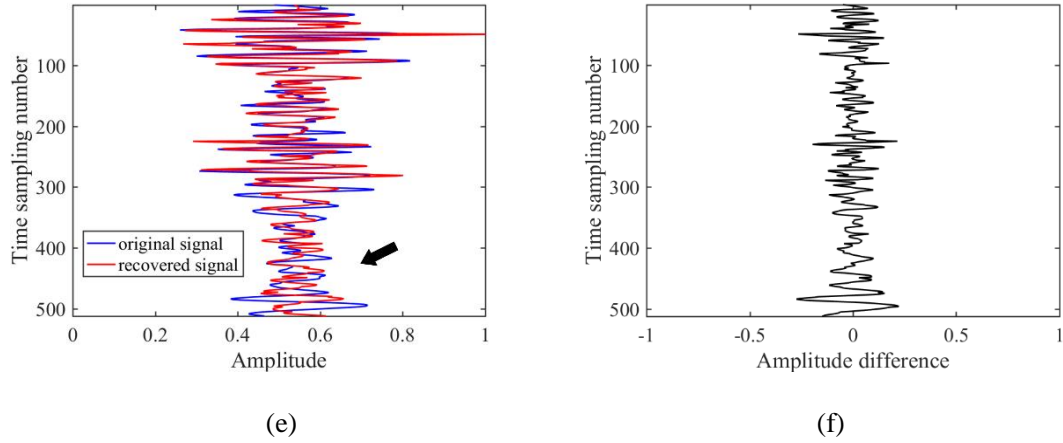

(e) (f)

Figure 11. Single-trace reconstruction of the 270th trace of the post-stack data, where blue line represents the complete original data and the red line represents the interpolated data: (a) and (b) results and residuals of the trained network $F$ in the SUIT algorithm; (c) and (d) results and residuals of the trained U-net model; and (e) and (f) results and residuals of the LMaFit method.

To verify the performance of the three methods under different rates of missing data, Table 1 lists the SNR of interpolation associated with the three methods with rates of missing data from 10% to 50%. At missing rates of 30% and 50%, the SNR of the reconstruction by the SUIT algorithm is 3.90 and 3.54 dB higher than U-net and 12.30 and 9.70 dB higher than LMaFit, respectively.

Table 1 The SNR of interpolation by the three studied methods for missing rates from 10% to 50%

|        | 10%   | 20%   | 30%   | 40%   | 50%   |
|--------|-------|-------|-------|-------|-------|
| Miss   | 10.85 | 7.50  | 5.45  | 4.57  | 3.00  |
| SUIT   | 48.12 | 44.83 | 41.09 | 36.81 | 33.47 |
| U-net  | 44.81 | 40.96 | 37.19 | 32.81 | 29.93 |
| LMaFit | 34.68 | 32.37 | 28.79 | 26.54 | 23.77 |

**3.3 Pre-stack data interpolation**

We further validated the algorithm performance on the pre-stack data. As the pre-stack data and the post-stack data are vastly different, the network trained on the post-stack data could not be used directly for interpolation of the pre-stack data, primarily because of the poor performance in the reconstruction results for events with steep dip angle. To make the network suitable for pre-stack data with such a high dip, we added a $512 \times 512$ size section of synthetic data to the training dataset, as shown in Fig. 12.

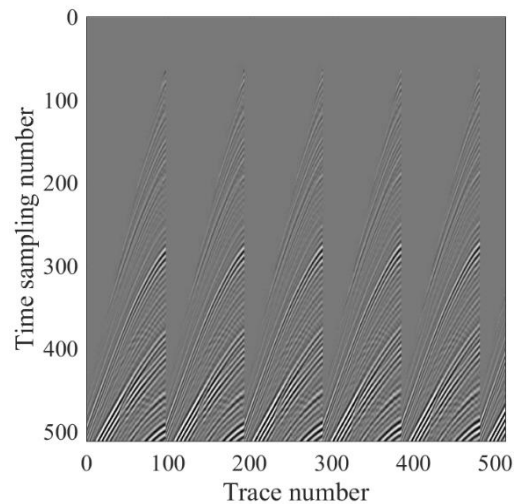

Figure 12. Pre-stack synthetic data

We used the trained network to interpolate the field pre-stack seismic dataset (Keys and Foster, 1998). The total dataset included 1001 shots, each of which included 120 traces and 1500 temporal samples per trace. Figure 13 shows the result of first field shot gather. The SNR of the SUIT reconstruction, U-net reconstruction, and LMaFit reconstruction is 12.62, 6.25, and 3.62 dB, respectively. As highlighted by the dotted box, the SUIT method presents better interpolation results on real pre-stack data when the training dataset is different and limited, and is more generalized than the U-net method. In addition, the result of the U-net method has large errors when there are multiple continuous missing traces, while the result of the LMaFit method has poor continuity regarding the events of this field shot gather.

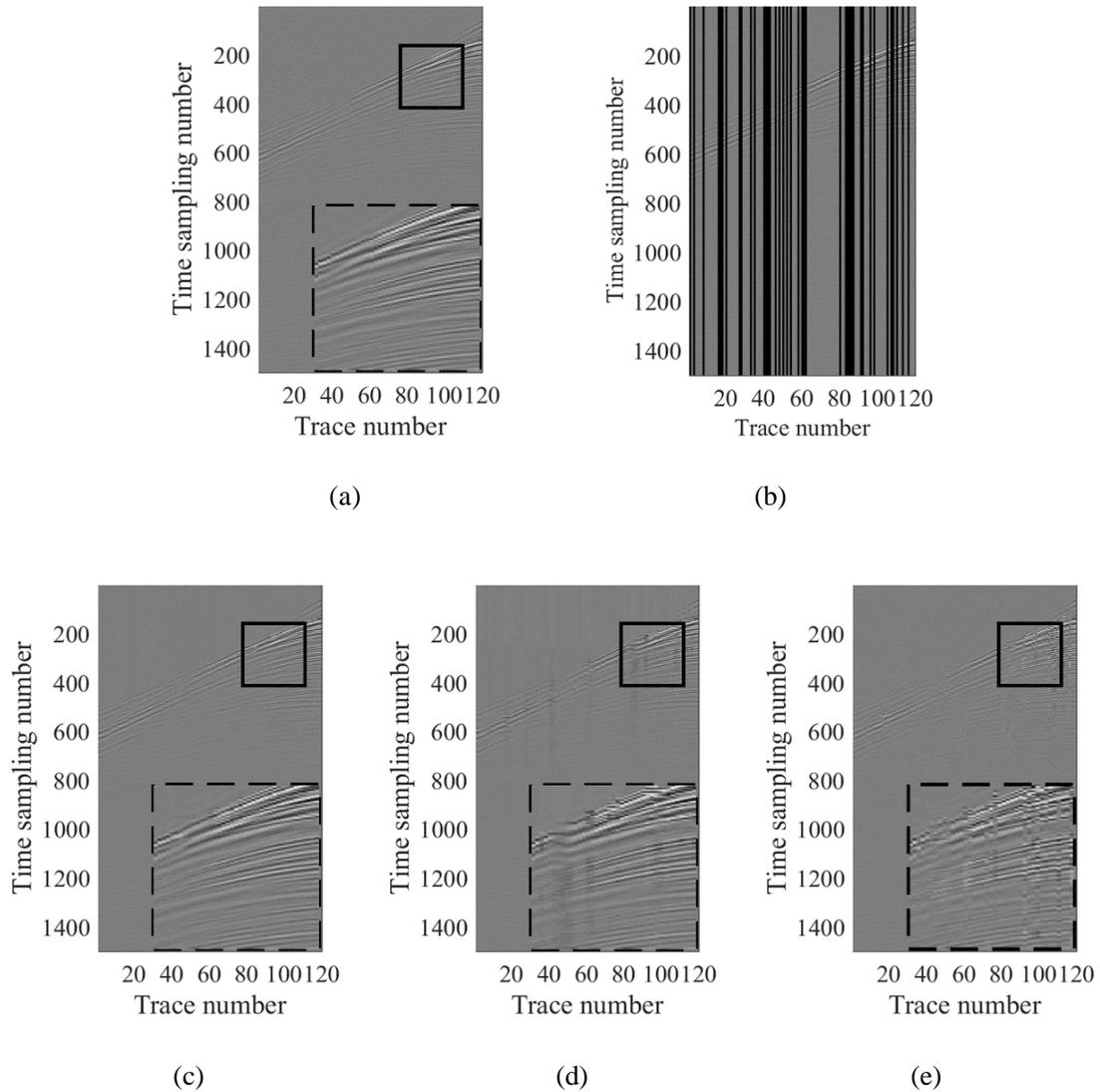

Figure 13. Comparisons of interpolation of pre-stack data: (a) complete data, (b) random sampled data with 30% missing, (c) interpolated data using the trained network $F$ in the SUIT algorithm, (d) interpolated data using the trained U-net model, and (e) interpolated data using the LMaFit method.

Figure 14 presents the *f-k* spectra of Figure 13. It can be seen that the reconstruction of the SUIT algorithm is consistent with the original and that it has better anti-aliasing capability, which further illustrates the validity of the SUIT algorithm

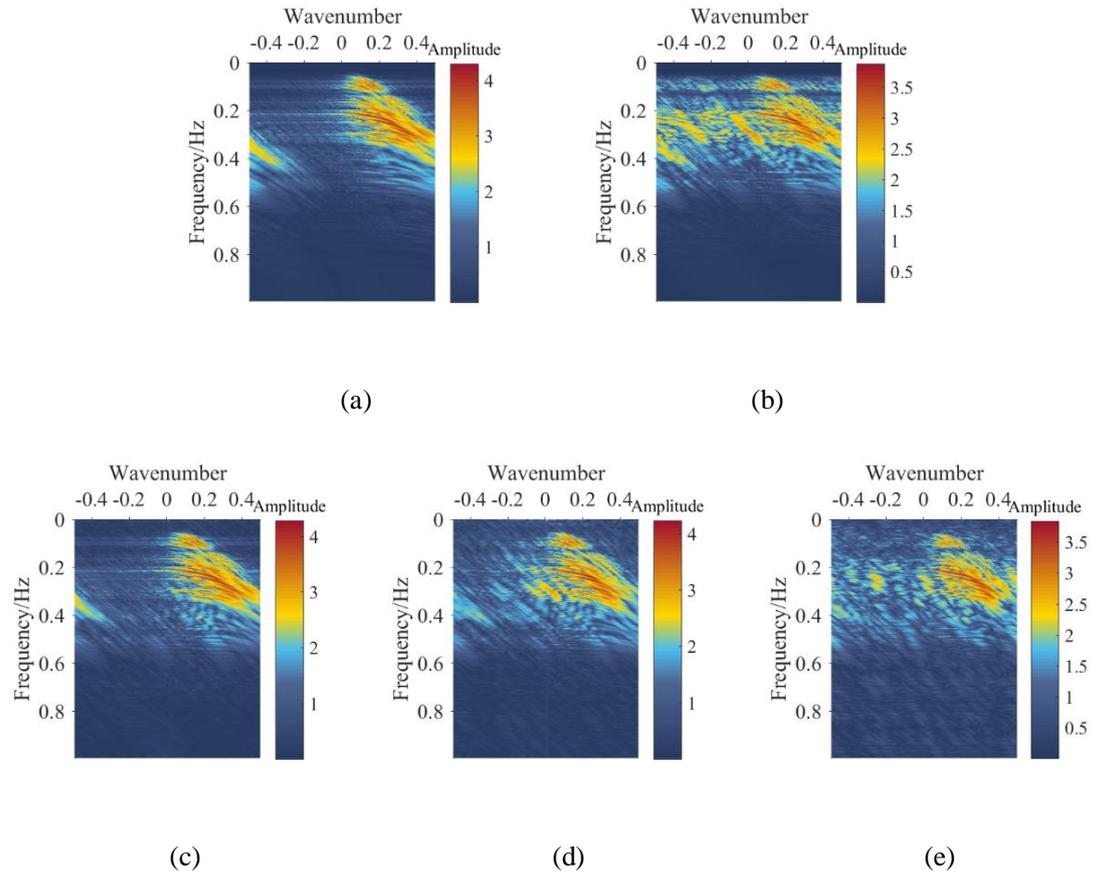

Figure 14. The *f-k* spectra of pre stack data; (a) complete data, (b) random sampled data with 30% data missing, (c) interpolated data using the trained network *F* in the SUIT algorithm, (d) interpolated data using the trained U-net model, and (e) interpolated data using the LMaFit method.

**3.4 Selection of parameter λ**

As shown in Equation (5), the loss function of the SUIT algorithm comprises two parts: reconstruction loss $L_r$ and texture loss $L_t$. Here, we discuss the impact of these two different losses on network training; the results of which are shown in Table 2. Taking $\lambda$ as 1, 50, 100, 150, and 200 successively, the average SNR of the verification dataset first increases and then decreases. The optimal value is approximately $\lambda = 100$. If the value of $\lambda$ is too small, the weight of $L_t$ becomes significant. Thus, for complex situations such as multiple continuous missing traces, it is difficult for the network to ensure the accuracy of the reconstruction texture. Consequently, to obtain smaller errors, it will fill the areas of missing data with noise instead, leading to a low SNR. When the value of $\lambda$ is too large, the weight of $L_t$ becomes small and the network will degenerate into an ordinary U-net model.

Table 2 SNR of the verification dataset under different $\lambda$

| $\lambda$ | 1 | 50 | 100 | 150 | 200 |
|---|---|---|---|---|---|
| SNR | 23.85 dB | 26.75 dB | 27.76 dB | 27.31 dB | 27.03 dB |

**4 Conclusion**

In recognition of the rich texture information of seismic data, we combined a basic U-net network with texture constraints to propose our SUIT algorithm. This algorithm was demonstrated to be capable of improving the accuracy and continuity of interpolation events by strengthening the learning of texture, and was shown to increase the reconstructed SNR. The SUIT algorithm first used the K-means algorithm to extract the texture of the seismic data. This was followed by the extraction of texture as labels with which to train a texture extractor and provision of the texture loss. Finally, the reconstruction network was optimized via reconstruction loss and texture loss. Owing to the consideration of texture information, the reconstruction network trained by the SUIT method on a small training dataset can also produce improved generalization.

Generalization has always been a huge challenge regarding DL-based methods for seismic data interpolation. In our experiment, slope was found to be an important factor affecting the generalization capability of the proposed method. The network trained on post-stack data with a gentle dip performed poorly for pre-stack data with a steep dip angle. We overcame this problem by adding a synthetic dataset with steep reflectors in the training dataset. In addition, the texture information considered in our work was obtained using the K-means algorithm, which generates approximate results. The use of a texture extraction method that is more accurate might further improve the reconstruction of the proposed SUIT algorithm.

**Acknowledgment**

This study is financially supported by the National Key R&D Program of China (No. 2018YFC1503705), Science and Technology Research Project of Hubei Provincial Department of Education (B2017597), Hubei Subsurface Multi-scale Imaging Key Laboratory (China University of Geosciences) under grants (SMIL-2018-06) and the Fundamental Research Funds for the Central Universities under Grants CCNU19TS020.